\documentclass{iopart}
\usepackage{iopams}

\usepackage[english]{babel}

\usepackage{amsfonts}
\usepackage{amssymb}
\usepackage{amscd}
\usepackage{graphicx}

\usepackage[latin1]{inputenc}

\newcommand{\eqref}[1]{(\ref{#1})}

\newcommand{\be}[1]{\begin{equation}\label{#1}}
\newcommand{\ee}{\end{equation}}
\newcommand{\ba}[1]{\begin{eqnarray}\label{#1}}
\newcommand{\ea}{\end{eqnarray}}

\newcommand{\diag}{\mbox{\rm diag}}

\begin{document}
\title[Multichannel generalization of EPP SUSY transformations]%
{Multichannel generalization of eigen-phase preserving supersymmetric transformations}

\centerline{\today}

\author{Andrey M Pupasov-Maksimov $^{1,2}$}

\address{$^1$ Universidade Federal de Juiz de Fora,
Juiz de Fora, MG,  Brazil}

\address{$^2$ OOO Expert Energo, Moscow, Russia}

\eads{\mailto{pupasov@phys.tsu.ru}}

\begin{abstract}
We generalize eigen-phase preserving (EPP) supersymmetric (SUSY) transformations
to $N>2$ channel  Schr\"odinger equation with equal thresholds. It is established
that EPP SUSY transformations exist only in the case of even number of channels, $N=2M$.
A single EPP SUSY transformation provides an $M(M-1)+2$ parametric deformation of the
matrix Hamiltonian without affecting eigen-phase shifts of the scattering matrix.
\end{abstract}

\section{Introduction}

In this paper we study $N$ channel radial Schr\"{o}dinger equation with equal thresholds.
Such equation may describe scattering of particles with internal structure, for instance, spin \cite{taylor:72,sparenberg:08,pupasov:11}.
Supersymmetric (SUSY) transformations allow analytical studies of Schr\"{o}dinger equation
with a wide class of interaction potentials \cite{Matveev-Salle,cannata:93,samsonov:07,sparenberg:08}.
In particular, the inverse scattering problem \cite{IS1,IS2} for two-channel Schr\"{o}dinger equation with equal thresholds
may be treated by combined usage of single channel SUSY transformations \cite{BSpJPA}
and eigen-phase preserving (EPP) supersymmetric (SUSY) transformations \cite{pupasov:10}.
These transformations conserve the eigenvalues 
 of the scattering matrix and modify its eigenvectors (coupling between channels),
 in a contrast 
to phase-equivalent SUSY transformations which do not modify scattering matrix at
all \cite{baye:87,sparenberg:96,sparenberg:97,samsonov:02}.

In \cite{pupasov:11} the two-channel neutron-proton potential was reproduced by
a chain of SUSY transformations, where the coupled channel inverse
scattering problem were decomposed
into the fitting of the channel phase shifts \cite{BSpJPA,samsonov:03} and the fitting of the mixing
between channels.
The fitting of the mixing
between channels was provided by the EPP SUSY transformations.

This paper extends the two-channel EPP SUSY transformations
to higher number of channels. 

The paper is organized as follows.
We first fix our notations and recall basics of SUSY transformations \cite{samsonov:07,amado:88a,amado:90,Leeb}. Given the explicit form
of a second-order SUSY transformation operator we study the physical sector of
SUSY transformations between
real and symmetric Hamiltonians.
We analyze the most general form of
a second-order SUSY transformation for the case of mutually
conjugated factorization energies.
Then we discuss applications 
of SUSY transformations to the scattering problems and
calculate how $S$-matrix transforms.

We start section \ref{sec:EPP-SUSY} re-examining the conservation of
its eigenvalues in the two-channel case.
There is the following asymptotic condition 
for EPP SUSY transformations.
The term with first-order derivative 
in the operator of
EPP SUSY transformation 
vanishes 
at large distances. 

To generalize EPP SUSY for arbitrary number of channels we study 
the matrix equation which comes from this asymptotic 
condition.
We find that EPP SUSY transformations may exist for
$2M$-channels only and obtain their general form.
A 4-channel example explicitly shows how EPP SUSY transformations acts.
We conclude with a summary of the obtained results and discussions of possible applications.

\section{Second order SUSY transformations}
\subsection{Definition of SUSY transformations}

SUSY transformations of stationary matrix Schr\"odinger equation are well known \cite{amado:88a,Leeb}.
In this subsection we just fix our notations.

Consider a family of matrix Hamiltonians $\mathbb{H}=\{H_a\}$
\begin{equation}\label{def-matrix-hamiltonian}
H_a=-I_N\frac{d^2}{dr^2}+V_a(r)\,,
\end{equation}
where $I_N$ is the $N\times N$ identity matrix, $V_a(r)$
is the
$N\times N$ real symmetric matrix potential. A multi-index $a$
parameterizes a family of potentials $V_a(r)$.
A matrix Hamiltonian defines the system of ordinary differential equations
\begin{equation}\label{def-schrodinger-equation}\fl
H_a\varphi_a(k,r)=k^2\varphi_a(k,r)\,.
\end{equation}
on $N\times N$ matrix functions $\varphi_a(k,r)$.

A (polinomial) SUSY transformation of \eqref{def-schrodinger-equation} is a
map of solutions
\begin{equation}\label{rel-mapping-solutions}
L_{ba}:\, \varphi_a(k,r)\to \varphi_b(k,r)=L_{ba}\varphi_a(k,r)\,,
\end{equation}
provided by a differential matrix operator
\begin{equation}\label{def-n-susy-transformation}
L_{ba} =A_n\frac{d^n}{dr^n}+A_{n-1}\frac{d^{n-1}}{dr^{n-1}}+\ldots+A_1\frac{d}{dr}+A_0\,,
\end{equation}
where $A_j$, $j=0,\ldots,n$, are some matrix valued functions.
This differential matrix operator obeys the intertwinning relation
\begin{equation}\label{rel-intertwinning}
L_{ba}H_a = H_bL_{ba} \,.
\end{equation}

The intertwinning relation \eqref{rel-intertwinning}
defines both the operator $L_{ab}$ and the transformed Hamiltonian $H_b$.
We will consider a family of Hamiltonians
$\mathbb{H}[H_a]=\{H_b| L_{ba}H_a = H_bL_{ba}\}$ related with the given
Hamiltonian $H_a$ by transformation operator \eqref{def-n-susy-transformation}.
In the next subsection we  present
explicit form of the second order transformation operator and the transformed Hamiltonian.

\subsection{Second-order SUSY algebra of matrix Schr\"odinger equation}
Given initial matrix Hamiltonian $H_0$, we choose two $N\times N$ matrix solutions
\begin{equation}\label{transformation-sol-def}
H_0u_j=E_ju_j\,, \qquad j=1,2\,,
\end{equation}
with $E_1$, $E_2$ called factorization constants. These functions
determine a second order operator $L_{20}[u_1,u_2]$:
\begin{equation}\label{Lu1u2-def}
\fl
L_{20}f(r)=\left[I_N\partial_r^2-V_0+E_1+(E_2-E_1)(w_1-w_2)^{-1}(w_1-\partial_r)\right]f(r)\,,
\end{equation}
where
\begin{equation}\label{def-superpotentials}
w_j(r)  =  u_j'(r)u_j^{-1}(r)\,,\qquad w_j^2+w_j'+E_j=V_0\,,\qquad j=1,2\,.
\end{equation}
Functions $w_j(r)$ are called superpotentials.
We also introduce a second-order superpotential
\begin{equation}\label{W2-def}
W_2(r) = (E_2-E_1)\left[w_1(r)-w_2(r)\right]^{-1}.
\end{equation}
The more symmetric and compact form of formula \eqref{Lu1u2-def} reads
\begin{equation}\label{Lu1u2-symmetric-def}
\fl
L_{20}f(r)=
\left[-H_0+
\frac{E_2+E_1}{2}+W_2\left(\frac{w_1+w_2}{2}-\partial_r\right)\right]f(r)\,.
\end{equation}

Operator $L_{20}$ and Hamiltonian $H_0$ obey the following algebra
\begin{equation}\label{intertwinning-relation}
L_{20}H_0=H_2L_{20}\,,\qquad H_2=H_0-2W_2'\,,
\end{equation}
\begin{equation}\label{factorization-property}
L_{20}^\dagger L_{20}=(H_0-E_1)(H_0-E_2)\,,\quad  L_{20} L_{20}^\dagger=(H_2-E_1)(H_2-E_2)\,.
\end{equation}

The new (transformed) potential is expressed in terms of second-order superpotential
$W_2$
as follows
\begin{equation}\label{transformed-potential}
V_2=V_0-2W_2'\,.
\end{equation}

The transformation operator has a global symmetry
\begin{equation}\label{transf-sol-global-symmetry}
L_{20}[u_1,u_2]=L_{20}[u_1(r) U_1,u_2 U_2]\,, \qquad {\rm det}\,U_{1,2}\neq 0\,.
\end{equation}

The second order SUSY transformations of an initial
Hamiltonian $H_0$ form the family $\mathbb{H}_2[H_0]=\{H_2|L_{20}H_0=H_2L_{20}\}$.
We will work only with Hamiltonians from $\mathbb{H}_2[H_0]$ and we omit
subscripts in the notation of transformation operator $L_{20}\rightarrow L$.

\subsection{Restrictions to the SUSY transformations}

We restrict our consideration only to second order SUSY transformations
with mutually conjugated factorization energies $E_1=E_2^*={\cal E}$.
Potentials  $V_0$ and $V_2$ are supposed to be
real and symmetric.
Hence the transformation functions
$u_1$ and $u_2$ have to be mutually conjugated, $u_1=u_2^*=u$.
The symmetry of $V_2$ demands the
symmetry of superpotentials \eqref{def-superpotentials},
$w_1^T=w_1=w$, $w_2^T=w_2=w^*$. Defining
the  Wronskian of two matrix functions as
\begin{eqnarray}\label{def-wronskian}
\mathrm{W}[u_1,u_2](r) & \equiv & u^T_1(r)u_2'(r)-{u^T_1}'(r)u_2(r) \\
    & = & u_1^T(r)\left[w_2(r)-w_1^T(r)\right] u_2(r). \label{Wronsk-spt}
\end{eqnarray}
we see that the symmetry of superpotential $w$
implies a vanishing self-Wronskian W$[u,u]=0$ of
transformation functions \cite{sparenberg:08}.

We present
the second-order superpotential $W_2$ in terms of the matrix Wronskian
for further needs,
\begin{equation}\label{def-twofold-superpotential-Wform}
W_2(r)  =  (E_1-E_2) u_2(r) {\mathrm W}[u_1,u_2]^{-1}(r)u_1^T(r)\,.
\end{equation}

To specify acceptable choice of transformation solutions explicitly,
we choose the basis in the solution space.
Natural basis for the radial problem, $r\in(0,\infty)$,
is formed by the Jost solutions $f(\pm k,r)$ with the exponential asymptotic
behavior
\begin{equation}\label{def-jost-solution}
f(k,r\rightarrow\infty)\to
I_Ne^{ikr}\,.
\end{equation}

Let us expand the transformation
functions in  the Jost basis
\begin{equation} \label{tf2}
u(r)=f_0(-K,r)C_j+f_0(K,r)D\,,
\end{equation}
where $K=k_r+i k_i$, $K^2={\cal E}$, $k_i>0$. Complex constant matrices $C$ and $D$
should provide vanishing self-wronskian W$[u,u]=0$.
The wronskian of two solutions with the same $k$
is a constant. For instance, ${\mathrm W}[f(-k,r),f(k,r)]=2ikI_N$.
Then, calculating W$[u,u]$ we get a constraint on the possible choice of matrices $C$ and $D$,
\begin{equation}\label{C-D-restriction}
D^TC=C^TD\,.
\end{equation}

Matrices $C$ and $D$
have an ambiguity due to symmetry \eqref{transf-sol-global-symmetry}.
Rank of matrix $C$, ${\rm rank}C = M\leq N$, determines the structure of 
transformation operator. The sum of ranks 
${\rm rank}C+{\rm rank}D \geq N$, otherwise operator
$L$ is undefined. Using \eqref{transf-sol-global-symmetry} we may
transform $C$ to the form, where only first $M$ columns are non-zero and linearly independent. Reordering
channels (by permutations of rows in the system of equations \eqref{def-schrodinger-equation}) we can put nontrivial $M\times M$ minor of $C$ into the upper left corner.
Then, $C$ and $D$ obey
the following canonical form,
\begin{equation}\label{C12}
C=\left(
\begin{array}{cc}I_{M}&0\\ Q &0\end{array}\right),
\qquad
D=\left(
\begin{array}{cc}X &-Q^T\\ 0 &I_{N-M}\end{array}\right),
\end{equation}
where $X=X^T$ is a symmetric $M\times M$ complex matrix,
and $Q$ is $(N-M)\times M$ complex  matrix.
This canonical form is a gauge which fixes ambiguity \eqref{transf-sol-global-symmetry}
of transformation solutions.

\subsection{Application to the scattering theory}

In concrete physical applications of SUSY transformations
we may further restrict the class of Hamiltonians.
In particular, in scattering theory \cite{taylor:72}
we work with the radial problem,
 $r\in(0,\infty)$. The interaction potentials
decrease sufficiently fast at large distances and
may contain centrifugal term
\begin{equation}\fl
\lim_{r\to\infty}r^2V(r)=l(l+I_N)\,,\qquad l={\rm diag}(l_1,\ldots,l_N)\,, \qquad {\rm e}^{il\pi}=\pm I_N\,.
\label{def-angular-momentum}
\end{equation}
The physical solution
has the following asymptotic behavior
\begin{equation}
\psi(k,r\to\infty) \propto k^{-1/2}
\left[ \rme^{-\rmi kr} \rme^{\rmi l\frac{\pi}{2}}
- \rme^{\rmi kr} \rme^{-\rmi l\frac{\pi}{2}} S(k) \right],
\label{phys-solution}
\end{equation}
where matrix coefficient $S(k)$ is the scattering matrix.

Scattering matrix is related with the Jost matrix
\begin{equation}\label{SdefI}
S(k)=e^{il\frac{\pi}{2}}F(-k)F^{-1}(k)e^{il\frac{\pi}{2}}\,,
\end{equation}
where the Jost matrix reads
\begin{equation}\label{Jost-matrix-def}
F(k)=\lim\limits_{r\rightarrow 0}
\left[f^T(k,r)r^\nu \right][(2\nu-1)!!]^{-1}\,.
\end{equation}
Diagonal matrix $\nu$ indicates the strength of the singularity
in the potential near the origin
\begin{equation}\label{Vzero}
V(r\rightarrow 0)=\nu(\nu+I_N)r^{-2}+{\rm O}(1)\,.
\end{equation}

Knowledge of the Jost solutions allows one to define scattering matrix.
Supersymmetric transformations of Hamiltonian and solutions
induce the transformation of scattering matrix.
Formal approach to the calculations of the S-matrices
was developed in the work of Amado \cite{amado:88b}.

Let us consider how the Jost solution transforms asymptotically,
\begin{eqnarray}\label{L-asymptot}
\fl
(Lf_0)(k,r\rightarrow\infty)=\\
\fl =\left[-k^2+
\frac{E_2+E_1}{2}+\left(W_2\frac{w_1+w_2}{2}\right)(r\rightarrow\infty)-ik W_2(r\rightarrow\infty)\right]\exp(ikr)\nonumber
\,.
\end{eqnarray}
Assume that there exists the following limit
\begin{equation}\label{def-Uinfty}
U_\infty(k)=\lim_{r\rightarrow\infty} \left[-k^2+
\frac{E_2+E_1}{2}+W_2\frac{w_1+w_2}{2}-ik W_2\right]\,.
\end{equation}
Then the transformed Jost solution reads
\begin{equation}\label{transformation-of-Jost-solution}
f_2(k,r)=(Lf_0)(k,r)U_\infty^{-1}(k)
\,,
\end{equation}

Making similar manipulations with the physical solution
\eqref{phys-solution} we establish the form
of transformed S-matrix
\begin{equation}\label{S-matrix-transformation}
S_2(k)=\rme^{\rmi l\frac{\pi}{2}} U_{\infty}(k)\rme^{-\rmi l\frac{\pi}{2}} S_0(k)\rme^{-\rmi l\frac{\pi}{2}}U^{-1}_{\infty}(k)\rme^{\rmi l\frac{\pi}{2}}\,.
\end{equation}

In the case of our second-order
SUSY transformation, the transformed S-matrix depends on the factorization energy ${\cal E}$
and parameters $Q$, $X$ through the matrix multipliers $U_{\infty}(k)$ and $U_{\infty}(k)^{-1}$.
In general, this dependence may be very complicated. Moreover, the scattering matrix $S_2$
may have unphysical low and high energy behavior.

SUSY transformations that deform the scattering matrix in a simple way are useful tools
to solve inverse scattering problem. In the two-cannel case there is a special kind
of deformation, when $U_{\infty}(k)$ becomes an orthogonal matrix \cite{pupasov:10}. We call such deformations as
eigen-phase preserving transformations.

\section{Eigen-phase preserving SUSY transformations \label{sec:EPP-SUSY}}
\subsection{Two channel case}

Let us analyze conditions that make a two-channel
SUSY transformation be an eigen-phase preserving one \cite{pupasov:10}.
In this case parameters of
the transformation, $Q=q$, $X=x$, are just some numbers.
Matrix $U_{\infty}(k)$ depends on $q$ only and becomes orthogonal
when $q=\pm i$. The determinant of $u$
vanishes at large distances,
 ${\rm det}\, u(r\rightarrow \infty)\rightarrow 0$ with such choice of $q$. Let
 ${\rm det}\, u(r\rightarrow \infty)\simeq \epsilon$, then superpotential
 $w$ diverges as $w(r\rightarrow \infty)\simeq \epsilon^{-1}$ and
 two-fold superpotential $W_2$ vanishes as $w_2(r\rightarrow \infty)\simeq \epsilon$.
As a result, the limit \eqref{def-Uinfty} contains only even powers of $k$
\begin{equation}\label{def-Uinfty-2channel}
U_\infty(k)=\lim_{r\rightarrow\infty} \left[-k^2+
\frac{E_2+E_1}{2}+W_2\frac{w_1+w_2}{2}\right]\,.
\end{equation}
The cancelation  of odd powers of $k$ is a necessary condition to provide
EPP SUSY transformations. In the next subsection we establish the most
 general form of matrix $Q$
 which leads to the vanishing limit
\begin{equation}\label{W2-limit-zero}
\lim_{r\rightarrow\infty} W_2=0\,,
\end{equation}
 for the case $N>2$.

Parameter $x$ is also
should be
fixed
to provide ${\rm det W}[u,u^*]\neq 0$ for all $r>0$ which leads to a finite $V_2$.

\subsection{Asymptotic SUSY transformation at large distances for arbitrary $N$}
The transformation function $u$  \eqref{tf2}
has the following
asymptotic behaviour at large distances
\begin{equation}\label{asymptotic-transformation-solution}
u(r\to\infty)\to u_\infty (I_N+ \Lambda r^{-1}+o(r^{-1}))\,,\qquad u_\infty=  A e^{-iKr \Sigma }\,,
\end{equation}
where
\begin{equation}\label{A-Sigma}
A=\left(
\begin{array}{cc}I_{M} &-Q^T\\ Q &I_{N-M}\end{array}\right)\,,\qquad
 \Sigma_{M,N-M}=\left(
\begin{array}{cc} I_{M}& 0\\ 0 & -
I_{N-M}\end{array}\right).
\end{equation}
For each concrete EPP transformation $N$ and $N-M$ are fixed,
therefore we will use notation $\Sigma$ instead of $\Sigma_{M,N-M}$.

The two-fold superpotential behaves asymptotically as
\begin{eqnarray}\label{spt2-asympt}
\lim_{r\to\infty} W_2&=& W_{2,\infty}  = ({\cal E}-{\cal E}^*) u_\infty^* {\mathrm W}[u_\infty,u_\infty^*]^{-1}u_\infty^T \label{spt2-wr-as} \\
    & = & 2i {\cal E}_{\rm Im}A^* e^{iK^*r \Sigma}{\mathrm W}[u_\infty,u_\infty^*]^{-1}e^{-iKr \Sigma}A^T.
\end{eqnarray}
Using asymptotic \eqref{asymptotic-transformation-solution} we see that this limit
is a constant matrix
\begin{eqnarray}\label{spt2-asympt-at}\fl
W_{2,\infty}     & = &
2i {\cal E}_{\rm Im}A^* e^{iK^*r \Sigma}\!\left[u_\infty^T \left(u_\infty^*\right)'-\left(u_\infty^T\right)' u_\infty^*\right]^{-1}e^{-iKr \Sigma}A^T\\
\fl \nonumber & = &
2i {\cal E}_{\rm Im}A^* e^{iK^*r \Sigma}\!\left[e^{-iKr \Sigma}\!A^T\!A^* \left(e^{iK^*r \Sigma}\right)\! '-\left(e^{-iKr \Sigma}\right)\! 'A^T\!A^* e^{iK^*r \Sigma}\right]^{\!-1}\!e^{-iKr \Sigma}A^T\\
\fl \nonumber & = &
2 {\cal E}_{\rm Im}A^*\left[K^* A^TA^*\Sigma +K \Sigma A^TA^* \right]^{-1}A^T\,.
\end{eqnarray}

We introduce auxiliary matrix  ${\rm W}_\infty$
\begin{equation}\label{Winfty}
\fl {\rm W}_\infty:=K^* A^TA^*\Sigma +K \Sigma A^TA^*=
2\left(
\begin{array}{cc}k_r(I_{M}+Q^TQ^*) & ik_i(Q^T-(Q^*)^T)\\ ik_i(Q-Q^*) & -k_r(QQ^\dagger+I_{N-M})\end{array}\right)\,.
\end{equation}
Limit \eqref{W2-limit-zero} leads to the following matrix equation
\begin{equation}\label{EPP-condition-W2zero}
W_{2,\infty}=0 \Rightarrow A^*{\rm W}_\infty^{-1}A^T=0\,, \qquad {\rm det W}_\infty \neq 0\,,
\end{equation}
which provides asymptotic cancelation of $k$ in \eqref{def-Uinfty}.
In the two channel case
these equations fix $Q$ uniquely. When $N>2$, these equations determine
a set of $Q$ values.

Equation \eqref{EPP-condition-W2zero}
may be satisfied if and only if matrix $A$ is singular.
Matrix ${\rm W}_\infty$ is invertible, ${\rm rank W}_\infty=N$.
Let ${\rm rank}\,A=n$, then ${\rm rank}\,A^*={\rm rank}\,A^T=n$
and ${\rm rank}({\rm W}_\infty^{-1} A^T)={\rm dim\,Img}({\rm W}_\infty^{-1} A^T)=n$.
The dimension of kernels ${\rm dim\, Ker}\,A={\rm dim\, Ker}\,A^*={\rm dim\, Ker}\,A^T=N-n$.
Equation \eqref{EPP-condition-W2zero} implies that
${\rm Img}({\rm W}_\infty^{-1} A^T)\subset {\rm Ker}\,A^*$, hence $n\leq N-n$.
Therefore equation \eqref{EPP-condition-W2zero} has solutions only if $n\leq \frac{1}{2}N$.
From the other hand, from explicit form of matrix $A$, \eqref{A-Sigma}, its rank
$n\geq {\rm max }(M,N-M)$. That is,
\eqref{EPP-condition-W2zero} has solutions if and only if
\begin{equation}\label{epp-rank-conditions}
{\rm rank}\,A=\frac{N}{2}\,,\qquad N=2M\,.
\end{equation}
From here it follows that for odd number of channels equation \eqref{EPP-condition-W2zero} has no solutions.

Consider  $2M\times 2M$ matrix $A$
\begin{equation}\label{A-2M}
A=\left(
\begin{array}{cc}I_{M} &-Q^T\\ Q &I_{M}\end{array}\right)\,,
\end{equation}
with ${\rm rank}\,A=M$. Two its rectangular sub matrices
have the same rank
\begin{equation}\label{A1-A2-2M}
{\rm rank}\left(
\begin{array}{c}I_{M} \\ Q \end{array}\right)={\rm rank}\left(
\begin{array}{c} -Q^T \\ I_M \end{array}\right)=M\,.
\end{equation}
We can take first $M$ columns of $A$ as
linearly independent, then from \eqref{epp-rank-conditions}, \eqref{A-2M} and \eqref{A1-A2-2M} follows that there exists $M\times M$
matrix $Z$, such that
\begin{equation}\label{Z-A1-A2}
\left(
\begin{array}{c}I_{M} \\ Q \end{array}\right)Z=\left(
\begin{array}{c} -Q^T \\ I_M \end{array}\right)=M\,.
\end{equation}
Solving this equation we obtain $Z=-Q^T$ and $QQ^T=-I_M$.

Let us extract $i$ from $Q$,
\begin{equation}\label{solution-for-Q}
Q=\pm i B\,,\qquad B^TB=BB^T=I_M\,,
\end{equation}
and substitute $Q$ in this form into \eqref{EPP-condition-W2zero}.
First of all we invert matrix ${\rm W}_\infty$. This matrix can be factorized
in two ways
\[
\fl
\frac{1}{2}{\rm W}_\infty
=\]
\[
\fl \left(
\begin{array}{cc}
k_rB^T(B+ B^*) & -k_i(B^T+B^\dagger)
\\
-k_i( B+ B^*) & -k_rB(B^\dagger+B^T)\end{array}\right)=
\left(
\begin{array}{cc}
k_r(B^\dagger+ B^T)B^* & -k_i(B^T+B^\dagger)
\\
-k_i( B+ B^*) & -k_r(B+B^*)B^\dagger\end{array}\right)=
\]
\[
\left(
\begin{array}{cc}
k_rB^T & -k_iI_M
\\
-k_iI_M & -k_rB\end{array}\right)
\left(
\begin{array}{cc}
(B+ B^*) & 0
\\
0 & (B^\dagger+B^T)\end{array}\right)=
\]
\[
\left(
\begin{array}{cc}
(B^\dagger+ B^T) & 0
\\
0 & (B+B^*)\end{array}\right)
\left(
\begin{array}{cc}
k_rB^* & -k_i I_M
\\
-k_iI_M & -k_rB^\dagger\end{array}\right).
\]
We note that
\begin{eqnarray}\label{Winfty-squared}
\fl
{\rm W}_\infty {\rm W}_\infty^*
=
4(k_r^2+k_i^2)
\left(
\begin{array}{cc}
(B^\dagger+ B^T) & 0
\\
0 & (B+B^*)\end{array}\right)
\left(
\begin{array}{cc}
(B+ B^*) & 0
\\
0 & (B^\dagger+B^T)\end{array}\right).
\end{eqnarray}
Therefore the inverse matrix reads
\begin{eqnarray}\label{Winfty-inversed}
\fl
{\rm W}_\infty^{-1}=
\frac{1}{2|K|^2}
\left(
\begin{array}{cc}
k_rB^\dagger & -k_i I_M
\\
-k_iI_M & -k_rB^*\end{array}\right)
\left(
\begin{array}{cc}
(B^T+B^\dagger)^{-1} &
0
\\
0 &
( B+ B^*)^{-1}\end{array}\right)=
\end{eqnarray}
\[
\frac{1}{2|K|^2}
\left(
\begin{array}{cc}
k_rB^\dagger(B^T+B^\dagger)^{-1} & -k_i( B+ B^*)^{-1}
\\
-k_i(B^T+B^\dagger)^{-1} & -k_rB^*( B+ B^*)^{-1}\end{array}\right)
\]

Let us introduce notations for auxiliary matrices
\[
\tilde B=
\left(
\begin{array}{cc}
(B^T+B^\dagger) &
0
\\
0 &
( B+ B^*)\end{array}\right),
\]
\[
B_k=
\left(
\begin{array}{cc}
k_rB^\dagger & -k_i I_M
\\
-k_iI_M & -k_rB^*\end{array}\right).
\]
Then $ W_{2,\infty}= {\cal E}_{\rm Im} A^* B_k \tilde B^{-1} A^T/(k_r^2+k_i^2)$.

Using the following matrix identities
\begin{equation}\label{BBB-commutators-1}
\fl B^T(B^T+B^\dagger)^{-1} =(B^*+B)^{-1}B^*\,,\qquad B^\dagger(B^T+B^\dagger)^{-1} =(B^*+B)^{-1}B\,,
\end{equation}
\begin{equation}\label{BBB-commutators-2}
\fl (B^T+B^\dagger)^{-1}B^T =B^*(B^*+B)^{-1}\,,\qquad (B^T+B^\dagger)^{-1}B^\dagger =B(B^*+B)^{-1}\,,
\end{equation}

\begin{equation}\label{BAT-comm}
\fl B^{-1} A^T=
\left(
\begin{array}{cc}
(B^T+B^\dagger)^{-1} &
0
\\
0 &
( B+ B^*)^{-1}\end{array}\right)
\left(
\begin{array}{cc}I_{M} & iB^T\\ -iB &I_{M}\end{array}\right)=
\end{equation}
\[
\fl \left(
\begin{array}{cc}I_{M} & iB^*\\ -iB^\dagger &I_{M}\end{array}\right)\left(
\begin{array}{cc}
(B^T+B^\dagger)^{-1} &
0
\\
0 &
( B+ B^*)^{-1}\end{array}\right),
\]

\begin{equation}\label{ABk}
\fl A^* B_k=
K^*\left(
\begin{array}{cc}
 B^\dagger &
-i I_M
\\
-iI_M
&
- B^*\end{array}\right),
\end{equation}
we see that
\[A^* B_k \tilde{B}^{-1}A^T=\]
\[\fl
K^*
\left(
\begin{array}{cc}
 B^\dagger &
-i I_M
\\
-iI_M
&
- B^*\end{array}\right)
\left(
\begin{array}{cc}I_{M} & iB^*\\ -iB^\dagger &I_{M}\end{array}\right)\left(
\begin{array}{cc}
(B^T+B^\dagger)^{-1} &
0
\\
0 &
( B+ B^*)^{-1}\end{array}\right)
=0\,.
\]
Thus \eqref{solution-for-Q} gives solutions of \eqref{EPP-condition-W2zero}.

Now we can calculate the  asymptotic form of a transformation operator explicitly.
To calculate asymptotic of $ W_{2}w$ we use the symmetry of
superpotential $w=u'u^{-1}=(u^T)^{-1}(u^T)'$,
\[\fl
\lim_{r\to\infty}W_{2}w=W_{2,\infty}w_\infty=
 2i{\cal E}_{\rm Im} u_\infty^* {\mathrm W}[u_\infty,u_\infty^*]^{-1}u_\infty^T
 (u_\infty^T)^{-1}(u_\infty^T)'=
 \]
 \[
 -2i K {\cal E}_{\rm Im}  A^*{\rm W}_\infty^{-1}\Sigma A^T=\frac{-i{\cal E}_{\rm Im}}{K^*} A^* B_k \tilde B^{-1}\Sigma A^T=
 \]
 \[
2 {\cal E}_{\rm Im} \left(
\begin{array}{cc}
-i B^\dagger &
 I_M
\\
-I_M
&
 -iB^*\end{array}\right)
 \left(
\begin{array}{cc}
(B^T+B^\dagger)^{-1} &
0
\\
0 &
( B+ B^*)^{-1}\end{array}\right)
\]

Matrix $W_{2,\infty}$ is real, therefore
\[
\Omega=W_{2,\infty}\frac{w_\infty+w^*_\infty}{2}={\rm Re} (W_{2,\infty}w_\infty)=
\]
\[
 2k_rk_i \left(
\begin{array}{cc}
i(B^T-B^\dagger) &
2 I_M
\\
-2I_M
&
 i(B-B^*)\end{array}\right)
 \left(
\begin{array}{cc}
(B^T+B^\dagger)^{-1} &
0
\\
0 &
( B+ B^*)^{-1}\end{array}\right)
\]
The matrix $U_{\infty}$ defined in \eqref{def-Uinfty} reads
\begin{equation}\label{Uinfty-explicit-form}
U_{\infty}(k^2)=\left(-k^2+k_r^2-k_i^2\right)I_N+\Omega\,,
\end{equation}

Matrix $\Omega$ is real, orthogonal (up to a normalization), $\Omega^T\Omega=4k_r^2k_i^2I_N$, and antisymmetric $\Omega=-\Omega^T$.
To establish its orthogonality
and antisymmetry one should use relations \eqref{BBB-commutators-1}, \eqref{BBB-commutators-2}.
With these two properties of $\Omega$ the matrix $U_{\infty}(k^2)$ becomes proportional to
the orthogonal matrix
\begin{equation}\label{Uinfty-orthogonality}
U_{\infty}(k^2)U_{\infty}^T(k^2)=\left((-k^2+k_r^2-k_i^2)^2+4k_r^2k_i^2\right)I_N
\end{equation}

That is the Jost solutions at large distances are rotated by orthogonal matrix $U_{\infty}$
\begin{equation}\label{Lu1u2-asymptotic-exp}
\fl
(Lf)(k,r\to\infty)\to
U_{\infty}\exp(ikr)\,,
\end{equation}

In this case the S-matrix transformation \eqref{S-matrix-transformation}
is just an energy-dependent orthogonal transformation,
\begin{equation}\label{S-matrix-epp-transformation}
S_2(k)=R_S(k^2) S_0(k)R_S^T(k^2)\,,
\end{equation}
with the orthogonal matrix, $R_S^TR_S=I_N$,
\begin{equation}\label{RS-matrix-epp-transformation}
 R_S=\rme^{\rmi l\frac{\pi}{2}} U_{\infty}\rme^{-\rmi l\frac{\pi}{2}}
 \left[(-k^2+k_r^2-k_i^2)^2+4k_r^2k_i^2\right]^{-1/2}\,.
\end{equation}
That is we obtain desired generalization of two-channel EPP SUSY transformations.

The above analysis is valid for an arbitrary $M\times M$ symmetric matrix $X$. Transformed
$S$-matrix $S_2$ depends on matrix $Q$ only. Therefore $X$ might provide additional $M(M+1)/2$ parametric
deformation of potential $V_2$ without affecting the S-matrix. From the other hand,
possibility of such deformations contradicts to the uniqueness of the inversion of the complete set of scattering data.
Therefore, there may exist only one matrix $X$ corresponding to one physical potential $V_2$. The EPP SUSY
transformation should be uniquely determined by the factorization energy, $M\times M$ complex orthogonal matrix $B$ and
a sign factor. In the next subsection we show how to fix matrix $X$ and prove that the corresponding potential $V_2$
is regular for all $r>0$.

\subsection{Eigen-phase preserving SUSY transformation near the origin}

To analyze the properties of EPP SUSY transformation in
the vicinity of $r=0$ we will use the solution
 \begin{equation}\label{rajs}
\varphi_0(k,r)=\frac{i}{2k}\left[f_0(-k,r)F_0(k)-f_0(k,r)F_0(-k)\right],
\end{equation}
vanishing at the origin
 \begin{equation}\label{rajs-bb}
\varphi_0(k,r\rightarrow 0)\to
{\diag}\left(\frac{r^{\nu_1+1}}{(2\nu_1+1)!!},\ldots,\frac{r^{\nu_N+1}}{(2\nu_N+1)!!}\right)\,,
\end{equation}
where $F_0(k)$ is the Jost matrix \eqref{Jost-matrix-def}.
We rewrite transformation solution in the basis $(\varphi_0(K,r),f_0(K,r))$
expressing $f_0(-K,r)$ from \eqref{rajs}
\begin{equation}\label{fm-in-f-phi}
f_0(-K,r)=\frac{2K}{i}\varphi_0(K,r)F_0^{-1}(K)+f_0(K,r)F_0(-K)F_0^{-1}(K),
\end{equation}
and substituting in \eqref{tf2}
\begin{equation} \label{ppt-tf-fphi-parametrization}
 u(r)=\frac{2K}{i}\varphi_0(K,r)F_0^{-1}(K)C
+f_0(K,r)(D+
s_0C)\,.
\end{equation}
where
\begin{equation} \label{small-scattering-matrix}
s_0=F_0(-K)F_0^{-1}(K)=\left(\begin{array}{cc} s_1 & s_2^T \\ s_2 & s_3 \end{array}\right),\qquad \rme^{\rmi l\frac{\pi}{2}}s_0\rme^{\rmi l\frac{\pi}{2}}=S_0\,.
\end{equation}
We can transform matrix $(D+
s_0C)$ to the form of matrix $D$ (without affecting $C$) multiplying $u$ from the right
\begin{equation}
(D+s_0C)\left(\begin{array}{cc} I_M & 0 \\ - (s_2\pm i s_3 B) & I_M \end{array}\right)=\left(\begin{array}{cc} \tilde{X} & \mp i B^T \\ 0 & I_M \end{array}\right),
\end{equation}
where
\begin{equation}
\tilde{X}=X+s_1\pm i(s_2^TB + B^Ts_2)-B^Ts_3B,
\end{equation}
Then the transformation  solution reads
\begin{equation} \label{ppt-tf-fphi-Xt}
\fl u(r)=\frac{2K}{i}\varphi_0(K,r)F_0^{-1}(K)\left(\begin{array}{cc} I_M & 0 \\ \pm i B & 0 \end{array}\right)
+f_0(K,r)\left(\begin{array}{cc} \tilde{X} & \mp i B^T \\ 0 & I_M \end{array}\right).
\end{equation}

Consider the case $\tilde{X}=0$. In this case the
potential $V_2$ is regular for all $r>0$. Let us prove this.
According to the Wronskian representation of the second-order
superpotential $W_2$ \eqref{Wronsk-spt}, the potential $V_2$ will be regular if and only if ${\rm det}{\rm W}[u,u^*](r)\neq 0$.

The derivative of the Wronskian ${\rm W}[u,u^*]$ reads
\begin{equation} \label{wr-der-ct}
{\rm W}[u,u^*]'(r)=({\cal E}-{\cal E}^*)u^T(r)u^*(r)\,.
\end{equation}
By construction W$[u,u^*]$ is an anti-Hermitian matrix, i.e.\
${\rm W}[u,u^*]=-{\rm W}^\dagger[u,u^*]$.
We represent transformation solution in a block-diagonal form
\begin{equation}\label{block-diag-form-of-transf-sol}
u(r)=\left(
\begin{array}{cc}
u_{11} & u_{12}\\
u_{21} & u_{22} \end{array}\right),
\end{equation}
with $M\times M$ matrix blocks. When $\tilde{X}=0$ these blocks obey the following
boundary conditions: $u_{11}(0)=0$, $u_{21}(0)=0$, $u_{12}(\infty)=0$, $u_{22}(\infty)=0$.
As a result the Wronskian
\begin{equation} \label{wronsk-block-form}
\fl
{\rm W}[u,u^*]=
\left(
\begin{array}{cc}
u_{11}^T & u_{21}^T\\
u_{12}^T & u_{22}^T \end{array}\right)
\left(
\begin{array}{cc}
u'{}_{11}^* & u'{}_{12}^*\\
u'{}_{21}^* & u'{}_{22}^* \end{array}\right)-
\left(
\begin{array}{cc}
u'{}_{11}^T & u'{}_{21}^T\\
u'{}_{12}^T & u'{}_{22}^T \end{array}\right)
\left(
\begin{array}{cc}
u_{11}^* & u_{12}^*\\
u_{21}^* & u_{22}^* \end{array}\right)
,
\end{equation}
has vanishing blocks at the origin, $\tilde W_{11}(0)=0$,
and at infinity $\tilde W_{22}(\infty)=0$.
Boundary behavior of $\tilde W_{12}$ is not determined.

Now we can calculate diagonal blocks of the Wronskian
integrating its
derivative
\begin{equation} \label{wronsk-block-form-derivative}
\fl
\frac{{\rm W}[u,u^*]'(r)}{({\cal E}-{\cal E}^*)}=
\left(
\begin{array}{cc}
\tilde W'_{11} & \tilde W'_{12}\\
\tilde W'{}_{12}^\dagger & \tilde W'_{22}
\end{array}\right)
=
\left(
\begin{array}{cc}
u_{11}^Tu_{11}^*+u_{21}^Tu_{21}^* & u_{11}^Tu_{12}^*+u_{21}^Tu_{22}^*\\
 u_{12}^Tu_{11}^*+u_{22}^Tu_{21}^* & u_{12}^Tu_{12}^*+u_{22}^Tu_{22}^*
\end{array}\right)\,.
\end{equation}
Integration of \eqref{wronsk-block-form-derivative} with the established boundary conditions
yields
\begin{equation}\label{wronsk-integrated-block-form}
\fl
{\rm W}[u,u^*](r)=
({\cal E}-{\cal E}^*)
\left(
\begin{array}{cc}
\int\limits_0^r\left(u_{11}^Tu_{11}^*+u_{21}^Tu_{21}^*\right)dt &
\tilde{W}_{12}\\
\tilde{W}_{12}^\dagger   &
-\int\limits_r^\infty\left(u_{12}^Tu_{12}^*+u_{22}^Tu_{22}^*\right)dt
\end{array}\right).
\end{equation}
Assume that there is a point $r_0$ where ${\rm det}{\rm W}[u,u^*](r_0)=0$.
Hence, matrix ${\rm W}[u,u^*](r_0)$ has at least one zero eigenvalue
\begin{equation}\label{zero-eigenvector-W}
{\rm W}[u,u^*](r_0)\vec{v}=0\,.
\end{equation}
Let us represent $N$ dimensional eigen-vector $\vec{v}$ as two $M$ dimensional vectors $\vec{v}_u$ and $\vec{v}_d$
and rewrite \eqref{zero-eigenvector-W} as a system of equations
\begin{eqnarray}\label{zero-eigenvector-W-system}
\tilde W_{11} \vec{v}_u + \tilde W_{12}\vec{v}_d=0\,,\\
\tilde W_{12}^\dagger \vec{v}_u+ \tilde W_{22} \vec{v}_d=0\,.
\end{eqnarray}
The first term of the scalar product
$(\vec{v}_u,\tilde W_{11} \vec{v}_u) +(\vec{v}_u, \tilde W_{12}\vec{v}_d)=0$, ($(\vec{a},\vec{b})=a_j^*b^j$) is positive
\[
(\vec{v}_u,\tilde W_{11} \vec{v}_u)=\]
\[
\fl(\vec{v}_u,\int\limits_0^{r_0}dt\left(u_{11}^Tu_{11}^*+u_{21}^Tu_{21}^*\right) \vec{v}_u)
=\int\limits_0^{r_0} \left( (u_{11}^* \vec{v}_u,u_{11}^*\vec{v}_u)+(u_{21}^* \vec{v}_u,u_{21}^* \vec{v}_u)\right)dt>0\,,
\]
therefore $(\vec{v}_u, \tilde W_{12}\vec{v}_d)=n_u<0$ is real and negative.
Now calculating scalar product $ (\vec{v}_d,\tilde W_{12}^\dagger \vec{v}_u)+(\vec{v}_d, \tilde W_{22} \vec{v}_u)=0$
with
negative second term
\[
(\vec{v}_d,\tilde W_{22} \vec{v}_d)=\]
\[
\fl-(\vec{v}_d,\int\limits_{r_0}^\infty dt\left(u_{12}^Tu_{12}^*+u_{22}^Tu_{22}^*\right) \vec{v}_d)
=-\int\limits_{r_0}^\infty \left( (u_{12}^* \vec{d}_u,u_{12}^*\vec{v}_d)+(u_{22}^* \vec{d}_u,u_{22}^* \vec{v}_d)\right)dt<0\,,
\]
we obtain  a contradiction,
$(\vec{v}_d, \tilde W_{12}^\dagger \vec{v}_u)=n_u^*=n_u>0$. This
contradiction proves that Wronskian ${\rm W}[u,u^*]$ have only non-zero eigenvalues for all $r>0$.
As a result ${\rm W}[u,u^*]$
is invertible, and hence both $W_2$ and $V_2$ are regular (finite) for all $r>0$.
Any non-zero $\tilde{X}$ will lead to the potential $V_2$ which is singular in some point $r_0$.

This prove completes our construction of multi-channel EPP SUSY transformations. In the next subsection we 
present an illustrative example.

\subsection{4-channel coupled potential}
\begin{figure}
\begin{center}
\scalebox{0.5}{\includegraphics{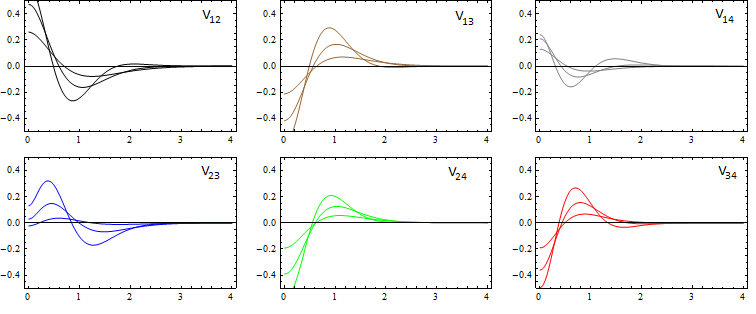}}
\caption{
Off-diagonal entries of the exactly solvable potential matrix $V_2$ obtained from the uncoupled potential \eqref{potential-4channel-example} with parameters
$a_1=1.1$, $a_2 = 1.5$, $a_3=2.1$, $a_4 = 2.5$, $b_r = 2.5$, $b_i = 1.3$, for 
three choices of the factorization energy
${\cal E}=-2+1.5 \rmi;-1.25+3.\rmi; 4.5 \rmi$. The strength of coupling increases with $\arg{\cal E}$ decreasing from
$0.78 \pi$ to $\pi/2$. 
\label{figV2ij}}
\end{center}
\end{figure}
\begin{figure}
\begin{center}
\scalebox{0.75}{\includegraphics{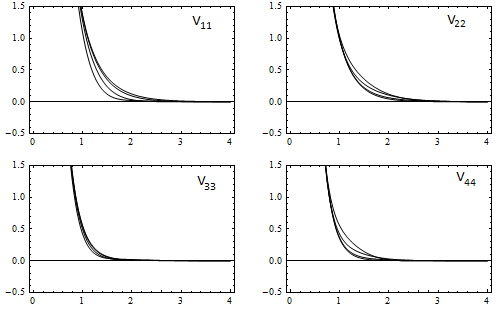}} \caption{
Diagonal entries of the exactly solvable potential matrix $V_2$ obtained from the uncoupled potential \eqref{potential-4channel-example} with parameters
$a_1=1.1$, $a_2 = 1.5$, $a_3=2.1$, $a_4 = 2.5$, $b_r = 2.5$, $b_i = 1.3$, for
three choices of the factorization energy
${\cal E}=-2+1.5 \rmi;-1.25+3.\rmi; 4.5 \rmi$.
\label{figV2ii}}
\end{center}
\end{figure}
We construct our initial 4-channel potential with $l=0$,
\begin{equation}\label{potential-4channel-example}
V_0(r)=\diag\left[ v_0(r,a_1),v_0(r,a_2),v_0(r,a_3),v_0(r,a_4)
\right],
\end{equation}
from four copies of
the following single channel potential
\begin{equation}\label{potential-1channel-example}
v_0(r,a)=\frac{2a^2}{\sinh^2(ar)}\,.
\end{equation}
Its scattering matrix is diagonal and reads
\begin{equation}\label{s-matrix-4channel example}
\fl S_0(k)=\diag\left[ s_0(k,a_1),s_0(k,a_2),s_0(k,a_3),s_0(k,a_4)
\right],\qquad s_0(k,a)=\frac{a-ik}{a+ik}\,.
\end{equation}
Consider an ingoing wave in $j$th channel
\begin{equation}\label{ingoing-wave-4channel-example}
\psi_{in,j}=\exp(-\rmi kr)(\delta_{1j},\delta_{3j},\delta_{3j},\delta_{4j})^T\,,
\end{equation}
where $\delta_{ij}$ is the Kroneker delta-symbol.
Ingoing wave is just a first term of
long-rage asymptotic
\begin{equation}
\left[ \rme^{-\rmi kr} 
- \rme^{\rmi kr} S_0(k) \right](\delta_{1j},\delta_{3j},\delta_{3j},\delta_{4j})^T\,.
\label{phys-solution-state-example}
\end{equation}
Scattering of such wave on the potential \eqref{potential-4channel-example} results just
in a phase shift
of the outgoing wave
\begin{equation}\label{outgoing-wave-4channel-example1}
\psi_{out,j}=\exp(\rmi kr+2 \rmi \delta_0(k,a_j))(\delta_{1j},\delta_{3j},\delta_{3j},\delta_{4j})^T\,,
\end{equation}
by eigen-phase
\begin{equation}\label{phase-shifts-1channel-example}
\delta_0(k,a_j)=-\arctan{\frac{k}{a_j}}\,,
\end{equation}
without mixing between channels.

Using EPP SUSY transformation we deform potential to introduce coupling between channels.
Diagonal components of the basis $(\varphi_0,f_0)$ explicitly reads
\begin{equation}\label{regular-solution-1channel-example}
\varphi_0(k,r;a)=\frac{1}{k^2+a^2}\left(k\cos(kr)-a \coth(ar)\sin(kr) \right)\,,
\end{equation}
\begin{equation}\label{jost-solution-1channel-example}
f_0(k,r;a)=\exp(ikr)\frac{k+ia\coth(ar)}{k+ia}\,.
\end{equation}
These solutions together with matrix $B$ depending on a single complex number
$b=b_r+ib_i$,
\begin{equation}\label{B-matrix-example}
B=\left(
\begin{array}{cc}
b & \sqrt{1-b^2}\\
-\sqrt{1-b^2} & b \end{array}\right),
\end{equation}
completely define EPP SUSY transformation.

Let us fix all parameters of the model $a_j, b_r, b_i$ except the factorization energy ${\cal E}$
($a_1=1.1$, $a_2 = 1.5$, $a_3=2.1$, $a_4 = 2.5$, $b_r = 2.5$, $b_i = 1.3$).
In Figures
\ref{figV2ij} and \ref{figV2ii} we show the potential $V_2$ provided
by EPP SUSY transformations for three values of ${\cal E}=-2+1.5 \rmi;-1.25+3.\rmi; 4.5 \rmi$.
The strength of coupling increases with $\arg{\cal E}$ decreasing from
$0.78 \pi$ to $\pi/2$. One can check that $\arg{\cal E}=0$ corresponds to zero-coupling, $V_2=V_0$. 
For our choice of matrix $B$ and parameters, the matrix $\Omega$ reads
\begin{equation}\label{Omega-example}
\fl \Omega={\cal E}_{\rm Im}\left(
\begin{array}{cccc}
 0 & -0.936848 & 0.305791 & -0.16973 \\
 0.936848 & 0 & 0.16973 & 0.305791 \\
 -0.305791 & -0.16973 & 0 & 0.936848 \\
 0.16973 & -0.305791 & -0.936848 & 0
\end{array}
\right)
\end{equation}

The matrix $S_2$ \eqref{S-matrix-epp-transformation} has the same eigenvalues, but non-diagonal character of potential results in the mixing of different channels in the outgoing wave,
\begin{equation}
\left[ \rme^{-\rmi kr} \rme^{\rmi l\frac{\pi}{2}}
- \rme^{\rmi kr} \rme^{-\rmi l\frac{\pi}{2}} R_SS_0(k)R_S^T \right](\delta_{1j},\delta_{3j},\delta_{3j},\delta_{4j})^T,
\label{phys-solution-state-example1}
\end{equation}

\begin{figure}
\begin{center}
\scalebox{0.75}{\includegraphics{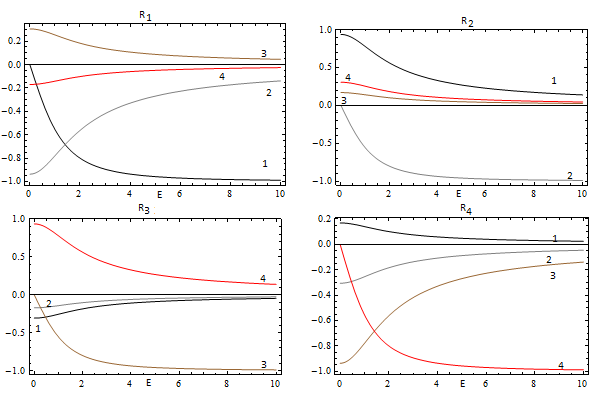}} \caption{
Eigenvectors of the scattering matrix $S_2$.
\label{figR}}
\end{center}
\end{figure}

There is another set of ingoing waves
\begin{equation}\label{mixed-ingoing-wave-4channel-example}
\psi_{in,j}=\exp(-\rmi kr)R_j(k^2)\,,\qquad R_S=(R_1,R_2,R_3,R_4)\,,
\end{equation}
given by columns $R_j(k^2)$ of matrix $R_S$ which scatter just with a phase shift \eqref{phase-shifts-1channel-example},
\begin{equation}\label{outgoing-wave-4channel-example2}
\psi_{out,j}=\exp(\rmi kr+2 \rmi \delta_0(k,a_j))R_j(k^2)\,.
\end{equation}
Vectors $\vec{R}_j$ depends on the energy of ingoing wave. In Figure \ref{figR} we show this dependence for a particular example 
${\cal E} = 4.5 \rmi$. Changing two complex parameters $b$ and \cal{E} we can manipulate 
transitions between channels which may open a way for broad physical application of EPP SUSY transformations.

\section{Conclusion \label{Concl}}

In the present paper we have generalized two-channel eigen-phase preserving SUSY transformations to the multichannel case,
$N=2M>2$. It was surprising, that such generalization exists for even number of channels only.
A single EPP SUSY transformation depends on a complex factorization energy ${\cal E}$, and
$M\times M$ complex matrix $B$, such that $B^TB=I_N$.
Therefore single EPP SUSY transformation provide an $M(M-1)+2$ parametric deformation of scattering matrix
without affecting eigen-phase shifts.

There are several possible applications of presented results. One can use  EPP SUSY transformations
to solve inverse scattering problem by deforming a diagonal S-matrix as in \cite{pupasov:11}.
We also may consider the S-matrix eigenvalues which conserved under $M(M-1)+2$ parametric deformation
as integrals of motions for some dynamical system associated with matrix Schr\"{o}dinger equation \cite{Matveev-Salle}.
In this context it is interesting to establish how this dynamical system looks. We expect that
in this way new exactly solvable non-linear equations may be discovered.

\section*{Acknowledgements}

AMP thanks Brazilian foundation CAPES (Coordena\c{c}\~{a}o de Aperfei\c{c}oamento de Pessoal de N\`{i}vel Superior) for the financial support.
This work is supported by RFBR grant N. 12-02-31552.


\section*{References}

\begin{thebibliography}{99}

\bibitem{taylor:72}
Taylor J R 1972 {\it Scattering Theory: The Quantum Theory on
  Nonrelativistic Collisions} (New York: Wiley)

\bibitem{sparenberg:08}
Sparenberg J-M, Pupasov A M, Samsonov B F and Baye D 2008
Exactly-solvable coupled-channel models from supersymmetric
quantum mechanics {\it Mod. Phys. Lett. B} {\bf 22} 2277-86

\bibitem{pupasov:11}
Pupasov A, Samsonov B~F, Sparenberg J~M and Baye D 2011 {\em Phys.\ Rev.\
  Lett.\/} {\bf 106} 152301


\bibitem{Matveev-Salle}Matveev V and Salle M 1991
{\it Darboux Transformations and Solitons} (New York: Springer)


\bibitem{cannata:93}
Cannata F and Ioffe M V 1993 Coupled channel scattering and
separation of coupled differential equations by generalized
Darboux transformations {\it J. Phys. A: Math. Gen} {\bf 26}
L89-92

\bibitem{samsonov:07}
Samsonov B F, Sparenberg J-M and Baye D 2007 Supersymmetric
transformations for coupled channels with threshold differences
{\it J. Phys. A: Math. Theor.}
 \textbf{40} 4225-40


\bibitem{pupasov:08b}
Pupasov A M, Samsonov B F and Sparenberg J-M 2008 Exactly-solvable
coupled-channel potential models of atom-atom magnetic Feshbach
resonances from supersymmetric quantum mechanics
{\it Phys. Rev. A} \textbf{77} 012724
({\it Preprint} quant-ph/0709.0343)



\bibitem{IS1}Levitan B M 1984
{\it Inverse Sturm-Liouville Problems} (Moscow: Nauka)

\bibitem{IS2}Chadan K and Sabatier P C 1989
{\it Inverse Problems in Quantum Scattering Theory}, 2nd
edn. (New York: Springer).

\bibitem{BSpJPA}Baye D and Sparenberg J-M 2004
{\it Inverse scattering with supersymmetric quantum
mechanics}
J. Phys. A: Math. Gen. {\bf 37} 10223-49


\bibitem{pupasov:10}
Pupasov A~M, Samsonov B~F, Sparenberg J~M and Baye D 2010 {\em J.\ Phys.\ A\/}
  {\bf 43} 155201


\bibitem{baye:87}
Baye D 1987
Supersymmetry between deep and shallow nucleus-nucleus potentials
{\it Phys. Rev. Lett.} {\bf 58} 2738-41

\bibitem{sparenberg:96}
Sparenberg J-M and Baye D 1996 Supersymmetry between deep and
shallow optical potentials for $^{16}$O + $^{16}$O scattering {\it
Phys. Rev. C} \textbf{54} 1309-21

\bibitem{sparenberg:97}
Sparenberg J-M and Baye D 1997
Supersymmetry between phase-equivalent coupled-channel potentials
{\it Phys. Rev. Lett.} \textbf{79} 3802-5




\bibitem{samsonov:02}
Samsonov B F and Stancu F 2002
Phase equivalent chains of Darboux transformations in scattering theory
{\it Phys. Rev. D} {\bf 66} 034001


\bibitem{samsonov:03}
Samsonov B F and Stancu F 2003 Phase shifts effective range
expansion from supersymmetric quantum mechanics {\it Phys. Rev. C}
\textbf{67} 054005

\bibitem{amado:88a}
Amado R D, Cannata F and Dedonder J-P 1988 Coupled-channel
supersymmetric quantum mechanics {\it Phys. Rev. A}  \textbf{38}
3797-800

\bibitem{amado:90}
Amado R D, Cannata F and Dedonder J-P 1990 Supersymmetric quantum
mechanics coupled channels scattering relations {\it Int. J. Mod.
Phys. A} \textbf{5} 3401-15



\bibitem{Leeb}
 Leeb H, Sofianos S A, Sparenberg J-M and Baye D 2000
 Supersymmetric transformations in coupled-channel systems
 {\it Phys. Rev. C} {\bf 62} 064003




\bibitem{amado:88b}
Amado R D, Cannata F and Dedonder J-P 1988 Formal scattering
theory approach to S-matrix relations in supersymmetric quantum
mechanics {\it Phys. Rev. Lett.} \textbf{61} 2901-4


\end{thebibliography}
\end{document}